\documentclass[12pt]{article}
\usepackage{graphicx}
\usepackage{epsfig,cite}
\newcommand{\be}{\begin{equation}}
\newcommand{\ee}{\end{equation}}
\newcommand{\beq}{\begin{equation}}
\newcommand{\eeq}{\end{equation}}

\newcommand{\bea}{\begin{eqnarray}}
\newcommand{\eea}{\end{eqnarray}}

\newcommand{\la}{\left\langle}
\newcommand{\ra}{\right\rangle}

\newcommand{\mrexp}{\mathrm{exp}}
\newcommand{\dat}{\mathrm{dat}}
\newcommand{\art}{\mathrm{art}} 
\newcommand{\rep}{\mathrm{rep}}
\newcommand{\net}{\mathrm{net}}
\newcommand{\sys}{\mathrm{sys}}
\newcommand{\stat}{\mathrm{stat}}
\newcommand{\diag}{\mathrm{diag}}

\newcommand{\lc}{\left[}
\newcommand{\rc}{\right]}
\newcommand{\lp}{\left(}
\newcommand{\rp}{\right)}
\newcommand{\bc}{\begin{center}}
\newcommand{\ec}{\end{center}}
\def\epm#1#2{\hbox{${\lower1pt\hbox{$\scriptstyle +~#1$}}
\atop {\raise1pt\hbox{$\scriptstyle -~#2$}}$}}

\newcommand{\fd}{F_2(x,Q^2)}

\def\gsim{\mathrel{\rlap{\lower4pt\hbox{\hskip1pt$\sim$}}
    \raise1pt\hbox{$>$}}}         
\def\lsim{\mathrel{\rlap{\lower4pt\hbox{\hskip1pt$\sim$}}
    \raise1pt\hbox{$<$}}}         

\begin{document}
\pagestyle{empty}
 
\begin{flushright}
 
{\tt hep-ph/0501067} \\CERN-PH-TH/2004-254\\IFUM-819/FT
\\  UB--ECM--PF 04/18\\GeF/TH/15-04\\
DFTT-30/04

\end{flushright}

\begin{center}

\vspace*{.6cm}

 {\bf{\Large Unbiased determination of the proton  structure\\ function
 $F_2^p$ with faithful uncertainty estimation}} \\

\vspace*{1.3cm}

{\bf  The NNPDF Collaboration:\\
Luigi Del Debbio$^1$, Stefano
  Forte$^2$,\\ Jos\'e I. Latorre$^3$, Andrea Piccione$^{4,\>5}$ and Joan Rojo$^3$, }

\vspace{0.5cm}

~$^1$ Theory Division, CERN, \\
 CH-1211 Gen\`eve 23, Switzerland\\
\medskip
~$^2$ Dipartimento di Fisica, Universit\`a di Milano and\\ 
INFN, Sezione di Milano, Via Celoria 16, I-20133 Milano, Italy\\
\medskip

~$^3$ Departament d'Estructura i Constituents de la Mat\`eria, \\
Universitat de Barcelona, Diagonal 647, E-08028 Barcelona, Spain\\
\medskip

~$^4$ INFN Sezione di Genova,\\ 
via Dodecaneso 33, I-16146 Genova, Italy\\
\medskip

~$^5$ Dipartimento di Fisica Teorica, Universit\`a di Torino,\\ 
via P.~Giuria 1, I-10125 Torino,  Italy\\

\vspace*{1.5cm}
                                                                 
{\bf Abstract}

\end{center}
\noindent

We construct a parametrization of the deep-inelastic structure
function
of the proton $\fd$ based on all available experimental information
from charged lepton deep-inelastic scattering experiments.
The parametrization  effectively provides a bias-free determination
of the probability measure in
the
space of structure functions, which retains information on experimental errors
and
correlations. The result is obtained in the form of a Monte Carlo
sample of neural networks trained on an ensemble of replicas of the
experimental data. We discuss in detail the techniques required for
the construction of bias-free parameterizations of large amounts of
structure function data, in view of future applications to the
determination of parton distributions based on the same method.

\vfill 
\begin{flushleft} 
December 2004 
\end{flushleft}

\eject   

\setcounter{page}{1} \pagestyle{plain}

\section{Introduction}
The requirements of precision physics at hadron colliders have recently led
to a rapid improvement in the techniques for the determination of
parton distributions of the nucleon, 
which are mostly extracted from deep-inelastic
structure functions~\cite{pdfs}. Specifically, it is now mandatory to
determine accurately the uncertainty on these quantities. The main
problem to be faced here is that one is trying to determine an
uncertainty on a function, i.e., a probability measure on a space of
functions, and to extract it from a finite set of experimental data.
This problem can be studied in a simpler context, namely, the
determination from the
pertinent data
of a structure function and its associate error. This sidesteps the
technical 
complication of
extracting parton distributions from structure functions, but it does
tackle the main issue, namely the determination of an error on a
function.
Furthermore, the determination of a structure function and associate
error might be useful for a variety of applications, such as precision
tests of QCD (determination of $\alpha_s$~\cite{netalpha}, tests of
sum rules) or the determination of polarized structure functions from
asymmetry data~\cite{smc}

A  new approach to this problem was recently proposed in
Ref.~\cite{nn}, based on the use of
neural networks as basic interpolating tools. The main idea of this
approach is to train a set of neural networks on a set of Monte Carlo
replicas of the experimental data which reproduces their probability
distribution. Hence, whereas the Monte Carlo replicas reproduce
faithfully the probability measure of the data for $F_2(x,Q^2)$ in the points of the
$(x,Q^2)$ plane where
data are available, the neural networks provide an interpolation and
extrapolation for all $(x,Q^2)$ subject to the only requirement of smoothness.
The set of
neural networks thus provides the desired probability measure, at
least in the measured $(x,Q^2)$ region, provided the sampling of the
$(x,Q^2)$ plane  is not too coarse.

In ref.~\cite{nn} a parametrization of the proton, deuteron and
nonsinglet $F_2$ structure functions based on the BCDMS and NMC fixed-target
deep-inelastic scattering data was constructed in this way. Here, we
extend the results of ref.~\cite{nn} by constructing a parametrization
of the proton $F_2$ structure function which includes all available
data, in particular the HERA collider data. Besides the obvious
motivation of having state-of-the art results for this quantity, the
main aim of this work is to develop a set of techniques which are
required for the application of the method of ref.~\cite{nn} to cases
where the handling of a large number of disparate data sets is
required. This involves in particular the use of genetic algorithms
for the training of neural networks.

In sect.~2 we summarize the features of the experimental data. In  
sect.~3 we review the  fitting method of ref.~\cite{nn}, emphasizing the
improvements which have been introduced here. In sect.~4 we discuss the
details of the training of neural nets to the current data set.
In sect.~5 our final results are presented and compared to those
previously obtained in ref.~\cite{nn}.

\section{Experimental data}

\begin{figure}[t]
\begin{center}
\epsfig{width=0.5\textwidth,figure=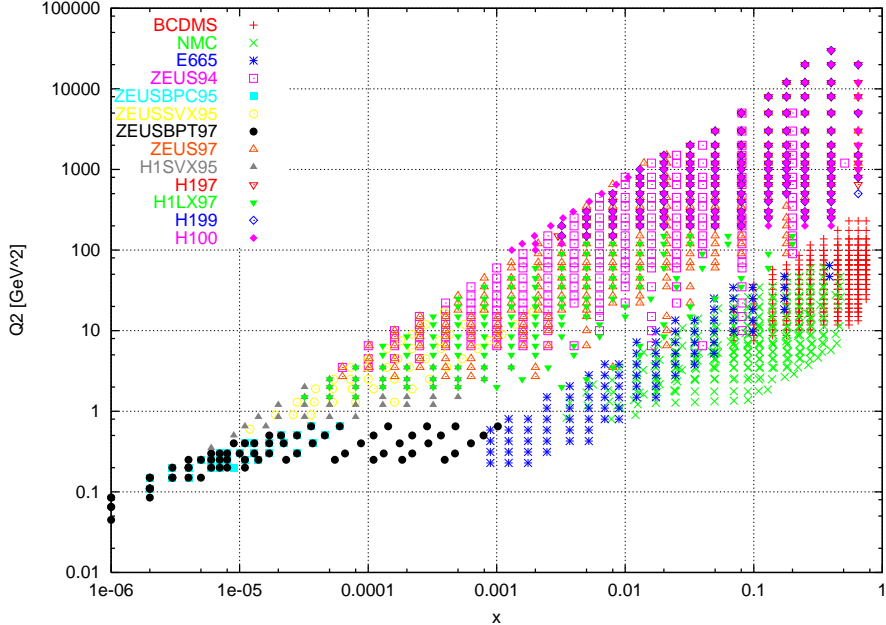,angle=-90}
\end{center}
\caption{\small Kinematic range of the experimental data}
\label{kinall}
\end{figure}
We construct a parametrization of $F_2$ based on all available unpolarized
charged lepton-proton deep-inelastic scattering
data~\cite{hepdata}. However, we do not include
early SLAC data, for which the covariance matrix is not available, since
they do not provide any extra kinematic coverage, and are anyway less precise
than later data.
This leaves a total of 13  experiments, listed in table~1, along with
their main features. The coverage of the $(x,Q^2)$ kinematic plane
afforded by these data is shown in fig.~1.

Structure functions are defined by parametrizing the deep-inelastic
neutral current scattering cross section as
\be
\frac{d^2\sigma(x,Q^2)}{dxdQ^2}=\frac{2\pi\alpha^2}{xQ^4}\lp
Y_+F_2(x,Q^2)\pm Y_-xF_3(x,Q^2)-y^2F_L(x,Q^2)\rp \ .
\label{f2}
\ee
For the definition of kinematic variables see ref.~\cite{pdg}.
We will construct a parametrization of the structure function
$F_2(x,Q^2)$, which provides the bulk of the contribution to
eq.~(\ref{f2}). For all experiments the $F_L$
contribution to the cross section  has already been subtracted by the
experimental collaborations, except for ZEUSBPC95, where
we subtracted it using the values published by the same experiment.
Note that the structure function $F_2$ receives contributions
from both $\gamma$ and $Z$ exchange, though the $Z$ contribution is
only non-negligible for the high $Q^2$ datasets
ZEUS94, H197,  H199 and H100. We will construct a parametrization of the structure
function $F_2$ defined in eq.~(\ref{f2}), i.e. containing all
contributions. When the experimental collaborations provide separately
the contributions to $F_2$ due to $\gamma$ or $Z$ exchange we have
recombined them in order to
get the full $F_2$ eq.~(\ref{f2}).

All the experiments included in our analysis provide full correlated
systematics, as well as normalization errors. 
The covariance matrix can be computed from these as 
\be
\label{covmat}
{\rm cov}_{ij}=
\lp\sum_{k=1}^{N_{\sys}}\sigma_{i,k}\sigma_{j,k}+F_iF_j\sigma_N^2\rp+\delta_{ij}\sigma^2_{i,t} \ ,
\ee
where $F_i$, $F_j$ are central experimental values, $\sigma_{i,k}$
are the
$N_{\sys}$ correlated systematics,
$\sigma_N$ is the total normalization uncertainty,
and the uncorrelated uncertainty  $\sigma_{i,t}$ is the sum of the
statistical uncertainty $\sigma_{i,s}$ and the $N_u$ uncorrelated
systematic uncertainties (when present):
\beq
\sigma_{i,t}^2= \sigma_{i,s}^2+\sum_{k=1}^{N_{u}} \sigma_{i,k}^2. 
\label{totuncorr}
\eeq
The correlation matrix
is then given by
\be
\rho_{ij}=\frac{ {\rm cov}_{ij}}{\sigma_{i,tot}\sigma_{j,tot}}
\label{cormat}
\ee
where the total error $\sigma_{i,tot}$ for the $i$-th point is given by
\beq
\sigma_{i,tot}=\sqrt{\sigma_{i,t}^2+\sigma_{i,c}^2+ F_i^2\sigma_N^2},
\label{toterr}
\eeq
the total correlated uncertainty $\sigma_{i,c}$ is the sum of all correlated systematics
\beq
\sigma_{i,c}^2=\sum_{k=1}^{N_{\sys}} \sigma_{i,k}^2.
\label{totsyst}
\eeq
\begin{table}[t]  
\begin{center}  
\tiny
\begin{tabular}{|cc|cc|c|ccccc|cc|} 
\hline
 Experiment & Ref. & $x$ range & $Q^2$ range & $N_{\dat}$   
& $\la\sigma_{\stat}\ra$& $\la\sigma_{\sys}\ra$& 
$\la\frac{\sigma_{\sys}}{\sigma_{\stat}}\ra$ &
$\la \sigma_{N}\ra$   & $\la\sigma_{tot}\ra$ 
&$\la\rho\ra$& $\la\mathrm{cov}\ra$ 
 \\
\hline
NMC & \cite{nmc} &$2.0~10^{-3}~-~6.0~10^{-1}$ & $0.5-75$ & 
 288 & 3.7 & 2.3 & 0.76  &2.0 &5.0 & 0.17 
& 3.8 
\\
\hline  
BCDMS & \cite{bcdms} & $6.0~10^{-2}~-~8.0~10^{-1}$ & $7-260$ &  351 & 3.2 & 2.0 
& 0.56 &3.0 & 5.4 & 0.52 & 13.1 
\\
\hline
E665 & \cite{e665} &$8.0~10^{-4}~-~6.0~10^{-1}$ & $0.2-75$ &  91 & 8.7 & 5.2 & 0.67
&2.0 & 11.0 &0.21 &
   21.7  
 \\
\hline
ZEUS94 & \cite{ZEUS94} &$6.3~10^{-5}~-~5.0~10^{-1}$ & $3.5-5000$ &
 188 &7.9 & 3.5 & 1.04 &  2.0
& 10.2  &0.12 & 6.4  
\\
ZEUSBPC95 & \cite{ZEUSBPC95} & $2.0~10^{-6}~-~6.0~10^{-5}$ & $0.11-0.65$ & 34 
& 2.9 & 6.6 & 2.38 &2.0 & 7.6 & 0.61 & 34.1  
\\
ZEUSSVX95 & \cite{ZEUSSVX95} & $1.2~10^{-5}~-~1.9~10^{-3}$ & $0.6-17$ 
&  44 & 3.8& 5.7 & 1.53 &1.0 & 7.1  & 0.10 &  4.1  
\\
ZEUS97 & \cite{ZEUS97} & $6.0~10^{-5}~-~6.5~10^{-1}$ & $2.7-30000$ &  240 & 
5.0 & 3.1 & 0.93 & 3.0 & 6.7 & 0.29& 7.0 
\\ZEUSBPT97 & \cite{ZEUSBPT97} &$6.0~10^{-7}~-~1.3~10^{-3}$ & $0.045-0.65$ &  70 
& 2.6 & 3.6 & 1.40 & 1.8 & 4.9  & 0.41 & 8.8
\\
\hline
H1SVX95 & \cite{H1SVX95} &$6.0~10^{-6}~-~1.3~10^{-3}$ & $0.35-3.5$ &  44 &  
 6.7& 4.6& 0.74 & 3.0 & 8.9 &0.36& 28.1 
\\
H197 & \cite{H197} & $3.2~10^{-3}~-~6.5~10^{-1}$& $150-30000$ & 130 & 
 12.5 & 3.2 & 0.31 &1.5 & 13.3 & 0.06 & 10.9 
\\
H1LX97 & \cite{H1LX97} & $3.0~10^{-5}~-~2.0~10^{-1}$ & $1.5-150$ &  133 &
2.6&2.2& 0.87 &1.7 & 3.9 & 0.30 & 3.9 
\\
H199 & \cite{H199} &$2.0~10^{-3}~-~6.5~10^{-1}$ & $150-30000$ &  126 &
 14.7& 2.8 &0.24 &1.8 & 15.2 & 0.05 & 11.0 
\\
H100 & \cite{H100} & $1.3~10^{-3}~-~6.5~10^{-1}$ & $100-30000$ &  147 &
9.4 & 3.2 & 0.42 &1.8 & 10.4 & 0.09 & 8.6 
\\
\hline

\end{tabular}
\caption{\small Experiments included in this analysis. All values of
  $\sigma$ and cov are given as percentages.}
\label{exp}
\end{center} 
\end{table}
\normalsize

For the ZEUS94,
ZEUSSVX95 and ZEUSBPT97 experiments some uncertainties are
asymmetric. As well known~\cite{barlow,dagostini,dagostini1}, asymmetric errors
cannot be combined in a simple multigaussian framework, and in
particular they cannot be added to gaussian errors in
quadrature. In
the treatment of multigaussian errors, we will follow the approach of
ref.~\cite{dagostini,dagostini1},  which, on top of several theoretical advantages, 
is closest to the ZEUS error
analysis and thus adequate for a faithful reproduction of the ZEUS
data. 
In this approach, a data point with central value $x_0$ and
left and right  asymmetric
uncertainties $\sigma_R$ and $\sigma_L$ (not necessarily positive) 
is described by a symmetric gaussian distribution, centered
at 
\be
\la x \ra=x_0+\frac{\sigma_R-\sigma_L}{2}\label{xshift}
\ee
and with width
\be
\sigma_x^2=\Delta^2=\left(\frac{\sigma_R+\sigma_L}{2} \right)^2 .
\label{Delta}
\ee
The ensuing distribution can then be treated in the standard
gaussian way.

\section{Fitting technique}
\label{strategy}

The construction of a parametrization of $F_2(x,Q^2)$ according to the
method of ref.~\cite{nn} consists of two steps:  generation of a
set of Monte Carlo replica of the original data,  and  training of a
neural network to each replica. We summarize here the main features
of these two steps, and the improvements that we introduced over the methods of
ref.~\cite{nn}. 

The Monte Carlo replicas of the original experiment are generated as a
multigaussian distribution: each replica is given by a set of values
\be
\label{replicas}
F_i^{(\art)(k)}=\lp1+r_{N}^{(k)}\sigma_N\rp\lp F_i^{\rm (\mrexp)}+
 \sum_{p=1}^{N_{\sys}}r_{p}^{(k)}\sigma_{i,p}+r_{i}^{(k)}\sigma_{i,t}\rp
 \ ,
\ee
where $F_i^{\rm (exp)}$ is the $i$-th data point, we introduce an
 independent univariate
 gaussian random number $r^{(k)}$  for each independent error
 source, and the various errors are defined in eqs.~(\ref{totuncorr}-\ref{toterr}).

\begin{figure}[t]
\begin{center}
\epsfig{width=0.3\textwidth,figure=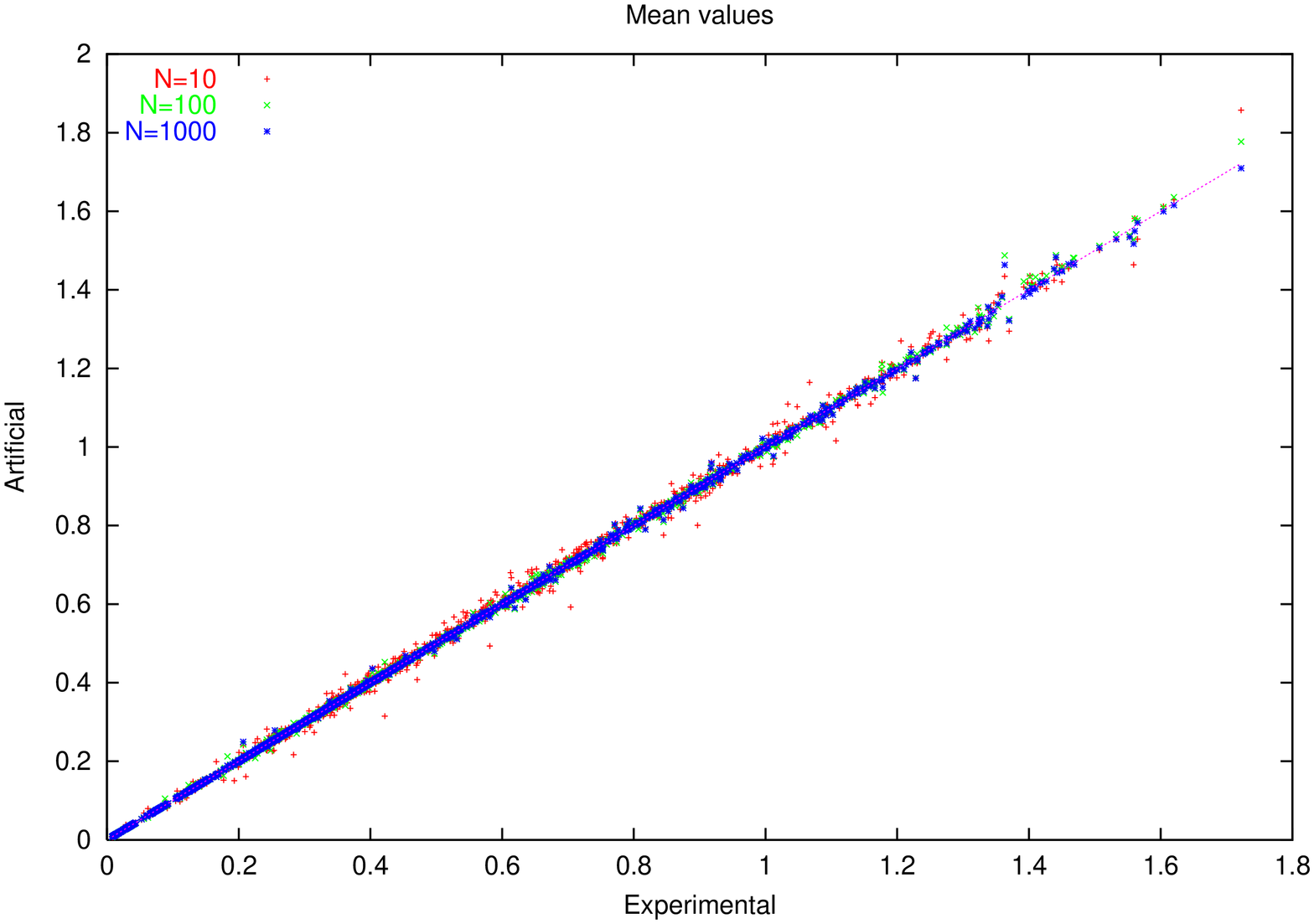,angle=-90}
\epsfig{width=0.3\textwidth,figure=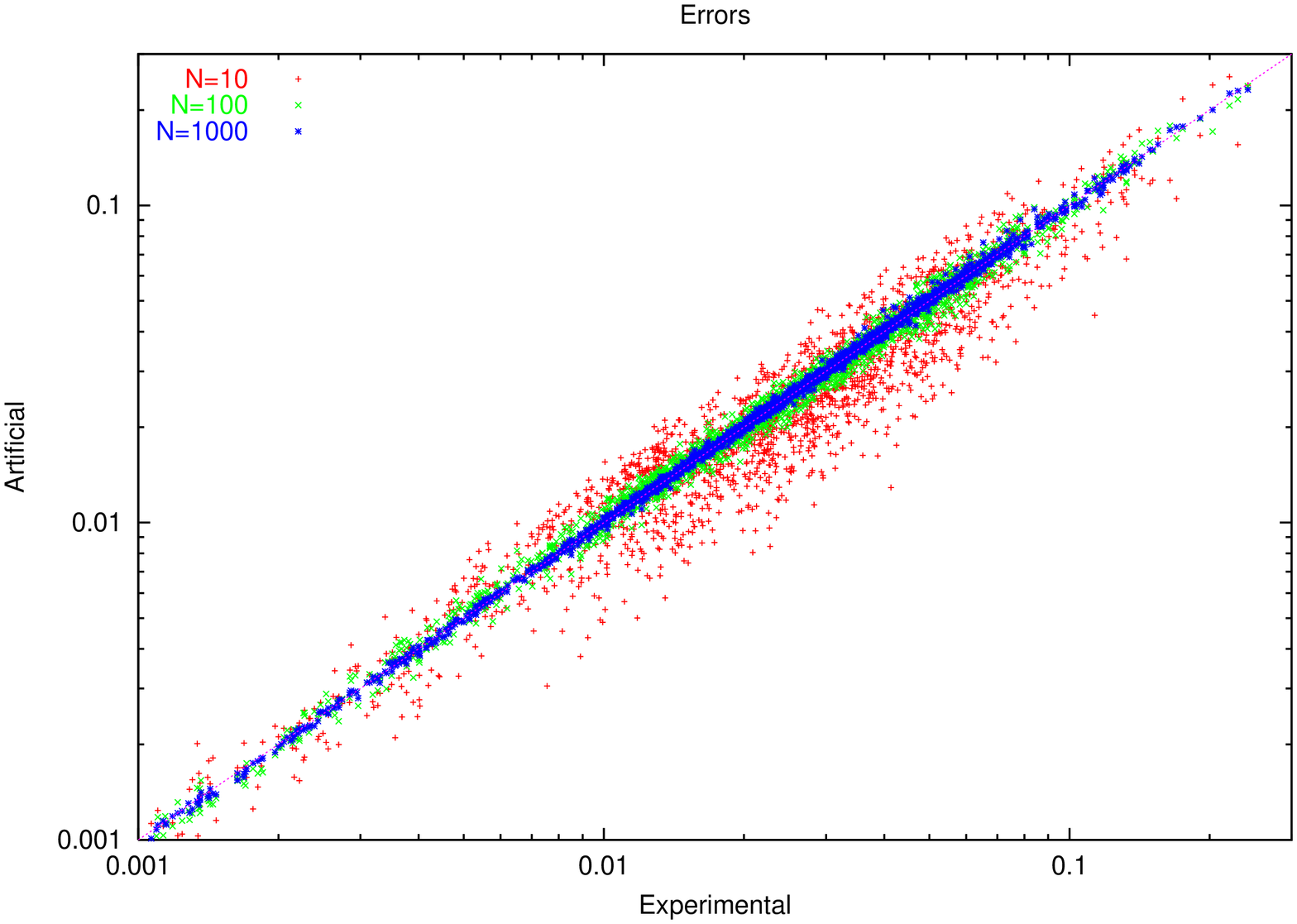,angle=-90}
\caption{\small
Scatter plot of experimental vs. Monte Carlo central values and errors.}
\label{genplots}
\end{center}
\end{figure}

The value of $N_{\rep}$ is determined in such a way that
the Monte Carlo set of replicas models faithfully the probability
distribution of the data in the  original  set.
A comparison of
expectation values, variance and correlation of the Monte Carlo set
with the corresponding input experimental values as a function of the
number of replicas is shown in fig.~\ref{genplots},
where we display scatter plots of the central values and errors 
for samples of 10, 100 and 1000 replicas. The corresponding plot for
correlations is essentially the same as  that shown in ref.~\cite{nn}.
A more quantitative comparison can be performed by defining suitable
statistical estimators (see the appendix). 
Results are presented in
table~\ref{gendata}. Note in particular the scatter correlations $r$
for central values, errors and correlations, which
indicate
the size of the spread of data around a straight line. The table~shows
that a sample of  1000 replicas is sufficient to ensure  average 
scatter correlations of 99\% and accuracies
of a few percent on structure functions, errors and correlations. 

\begin{table}[t]  
\begin{center}  
\begin{tabular}{|c|ccc|} 
\multicolumn{4}{c}{
$F_2^p(x,Q^2)$}\\   
\hline
 $N_{\rep}$ & 10 & 100 & 1000 \\
\hline
$\la PE\lc\la F^{(\art)}\ra_{\rep}\rc\ra$ & 1.88\% 
& 0.64\%  &  0.20\%    \\
$r\lc F^{(\art)} \rc$ & 0.99919 & 0.99992 &  0.99999 \\
\hline
$\la V\lc \sigma^{(\art)}\rc\ra_{\dat}$ & $6.7\times 10^{-4}$ &
$2.0\times 10^{-4}$ & $6.9\times 10^{-5}$  \\
$\la PE\lc \sigma^{(\art)}\rc\ra_{\dat}$ & 37.21\%
& 11.77\% & 3.43\%     \\

$\la \sigma^{(\art)}\ra_{\dat}$ & 0.0292 & 0.0317& 0.0316\\
$r\lc \sigma^{(\art)}\rc$ & 0.945 &  0.995 & 0.999 \\
\hline
$\la V\lc \rho^{(\art)}\rc\ra_{\dat}$ & $8.1\times 10^{-2}$  
&  $7.8\times 10^{-3}$ &  $7.3\times 10^{-4}$ \\

$\la \rho^{(\art)}\ra_{\dat}$ & 0.3048 & 0.3115 & 0.2920\\
$r\lc \rho^{(\art)}\rc$ & 0.696 &  0.951 & 0.995 \\
\hline
$\la V\lc \mathrm{cov}^{(\art)}\rc\ra_{\dat}$ & $5.2\times 10^{-7}$  
&  $6.8\times 10^{-8}$ &  $6.9\times 10^{-9}$ \\

$\la \mathrm{cov}^{(\art)}\ra_{\dat}$ & 0.00013 & 0.00018 & 0.00015\\
$r\lc \mathrm{cov}^{(\art)}\rc$ & 0.687 &  0.941 & 0.994 \\
\hline

\end{tabular}
\caption{\small Comparison between experimental and Monte Carlo data.\hfill\break
The experimental data have
$\la \sigma^{(\mrexp)}\ra_{\dat}=0.0311$, $\la \rho^{(\mrexp)}\ra_{\dat}=
0.2914$ and $\la \mathrm{cov}^{(\mrexp)}\ra_{\dat}=0.00015$. All
statistical indicators are defined in the appendix.}
\label{gendata}
\end{center} 
\end{table}

$N_{\rep}$ neural networks~\cite{netrev} are then trained on the Monte
Carlo data, by training each
neural network on all the $F_i^{(\art)}(k)$ data points in the
  $k$-th replica. The architecture of the networks is the same as in
  ref.~\cite{nn}. The training is
subdivided in three epochs, each
based on the minimization of a different error function. 
In the first training epoch, the networks are trained to minimize the function
\be
E_1^{(k)}=\frac{1}{N_{\dat}}\sum_{i=1}^{N_{\dat}}\lp F_{i}^{(\art)(k)}-
 F_{i}^{(\net)(k)}\rp^2 \ ,
\label{chinoerr}
\ee
i.e., the deviation from the central value per data point. In
the second epoch the function to be minimized is the uncorrelated
$\chi^2$ per data point, namely, the $\chi^2$ computed 
omitting correlated systematics: 
\be
{E_2^{(k)}=\chi^2_{\rm diag}}^{(k)}\equiv\frac{1}{N_{\dat}}
\sum_{i=1}^{N_{\dat}}\frac{\lp F_{i}^{(\art)(k)}-
 F_{i}^{(\net)(k)}\rp^2}{\sigma_{i,t}^{(\mrexp)^2}},
\label{chidiag}
\ee
where $\sigma_{i,t}^{(\mrexp)^2}$ is defined in eq.~(\ref{totuncorr}).
Finally, in
the third epoch the full $\chi^2$ per data point  is minimized, namely
\be
\label{chicov}
E_3^{(k)}
=\frac{1}{N_{\dat}}\sum_{i,j=1}^{N_{\dat}}
\lp
F_i^{(\art)(k)}-F_i^{(\net)(k)}\rp{\overline{\mathrm{cov}}^{(k)}}^{-1}_{ij}
\lp
F_j^{(\art)(k)}-F_j^{(\net)(k)}\rp \ ,
\ee
where ${\overline\mathrm{cov}^{(k)}}_{ij}$ is the covariance matrix
for the $k$-th replica, defined 
as in eq.~(\ref{covmat}), but with the normalization uncertainty
included as an overall rescaling of the error due to the normalization
offset of that replica:
namely, 
\be
\label{covmatnn}
{\overline\mathrm{cov}^{(k)}}_{ij}=
\lp\sum_{p=1}^{N_{\sys}}{\overline\sigma^{(k)}}_{i,p}{\overline\sigma^{(k)}}_{j,p}\rp+\delta_{ij}{\overline\sigma^{(k)}}_{i,t} \ ,
\ee
with 
\beq
\overline \sigma_{i,a}^{(k)}= (1+r^{(k)}_N\sigma_N)\sigma_{i,a} ,
\eeq
where $r^{(k)}$ is as in eq.~(\ref{replicas}). This is necessary in
order to avoid a biased treatment of the normalization
errors~\cite{dagostini1,dagostini2}. 

The rationale behind this three-step procedure is that the true minimum which
the fitting procedure must determine is that of the full $\chi^2$
eq.~(\ref{chicov}). However, this is nonlocal and
time consuming to compute. It is therefore advantageous to 
look for its rough features at first, 
then refine its search, and finally determine its actual location.

The minimum during the first two epochs is found using
back-propagation (BP) (see ref.~\cite{nn}). This method is  not
suitable for the minimization of the nonlocal function
eq.~(\ref{chicov}). 
In ref.~\cite{nn} BP was used throughout, and the
third epoch was omitted.  This is acceptable
provided the total systematics is small in comparison
to the statistical errors, and indeed it was verified that a
good approximation to the minimum of eq.~(\ref{chicov}) could be
obtained from the ensemble of neural networks. This is no longer the
case for the present extended data set, as we shall see explicitly in
section~5.  
Therefore, the full $E_3$ eq.~(\ref{chicov}) is
minimized in the third training
epoch by means of genetic algorithms (GA), previously used and discussed by
some of us for related purposes in ref.~\cite{jo}. 

In comparison to previous work~\cite{nn,jo}, we have implemented
several improvements, both in the BP and GA training epochs.
In the BP epoch, we use on line training as in ref.~\cite{nn},
i.e. the parameters of the neural network are updated after each data
point has been shown to it. This defines a training cycle.
In ref.~\cite{nn} it was shown that  the length of training needed to
achieve a given value of $E_3$ can differ significantly
between experiments:  it is larger for experiments
which have smaller errors,  contain more data, or  cover
kinematic regions where the structure function varies more rapidly. If one
wants to end up with a similar value of $E_3$ for all
experiments it is then
necessary to adjust the relative length of training of different data sets.
In ref.~\cite{nn} this was achieved by finding by trial and error
an optimal fixed weight for
the two experiments included in the fit. This procedure is
clearly not viable when the  number of experiments is large. 
Therefore, we have implemented a dynamical weighted
training, whereby the weight $P_i$ of each experiment is chosen initially to
be the same for all experiments, and then adjusted dynamically
according to the relative contribution of each experiment $E_{3,\,i}$
to the total $E_3$ eq.~(\ref{chicov}):
\be
P_i\to P_i \frac{E_{3,\,i}}{\sum_{j=1}^{N_{\exp}}E_{3,\,i}} .
\ee
The value of $E_{r,\,i}$ is updated  from the full data set every $1.25\times
10^6$ training cycles; because there are $\sim1700$ data points, this
ensures that between updates each data point has been seen by the net about 700
times on average in the unweighted case.
This method is not viable in the third (GA) training
epoch, where $E_3$ can only be computed from the full set of
data points (i.e., the training is necessarily batched, and not
on-line). Therefore, one cannot choose to
show a subset of data more often. One could in principle reweight the
contribution of the single experiments to $E_3$, but this
might distort the global minimum in an unpredictable way.

In the GA epoch, we have introduced two improvements in comparison to
the methods described in ref.~\cite{jo}. First, we have introduced
multiple mutations, specifically
three nested mutations for each training cycle.\footnote{Note that GA training cycles in  ref.~\cite{jo} are
referred to as
  generations, as it is customary for genetic algorithms.}  The purpose of this is to
avoid  local minima, thereby increasing the speed of 
training. It is crucial that  rates for these additional mutations are large,
in order to allow for jumps from a local minimum to
a deeper one. We find that
one additional mutation with probability 20\%, followed by 
 two additional mutations with probability 4\%,
produce a significant improvement of the convergence rate.
Second, we have introduced probabilistic selection. This entails that 
the sample of $N_{\mathrm{sel}}$ selected mutations is constructed by 
always selecting the mutation $i_0$ with the lowest
$E_3$ value, plus $N_{\mathrm{sel}}-1$ mutations selected among the total
$N_{\mathrm{mut}}$ mutations with
probability
\be
P_j=\exp\lp-\frac{E_{3,\,j}- E_{3,\,i_0}}
{E_{3,\,i_0}}\rp .
\ee
Namely, mutations with  larger $E_3$ values are less likely to be selected
but can still be selected with finite probability.
This is helpful in allowing for mutations which only become
beneficial after a combination of several
individual  mutations. 

At the end of the GA training we are left with a sample of $N_{\rep}$
neural networks, from which e.g. the value of the structure function
at $(x,Q^2)$ can be computed as
\beq
F_2(x,Q^2)=\frac{1}{N_{\rep}}\sum_{k=1}^{N_{\rep}} F^{(\net)(k)} (x,Q^2).
\label{avf2}
\eeq
The goodness of fit of the final set is thus measured by the $\chi^2$
per data point, which, given the large number of data points is
essentially identical to the $\chi^2$ per degree of freedom:
		\be
\chi^2=\frac{1}{N_{\dat}}\sum_{i,j=1}^{N_{\dat}}\lp \la
F_i^{(\net)}\ra_{\rep}-
F_i^{(\mrexp)}\rp \lp \mathrm{cov}^{-1}\rp_{ij} \lp\la
F_j^{(\net)}\ra_{\rep}-
F_j^{(\mrexp)}\rp \ ,
\label{chi2}
\ee
where the average over replicas, denoted by $\la\ra_{\rep}$, is defined in
the appendix.

\section{Training to the $F_2^p$ data}

In order to apply the general method discussed in sect.~3 to the
$F_2^p$ data presented in sect.~2 several specific issues must be
addressed: the choice
of training parameters and training length, the choice of the actual
data set, and the choice of theoretical constraints.
We now address these issues in turn.

The parameters and length
for the first two training epochs have been determined by
inspection of the fit of a neural network to the central experimental
values. Clearly, this choice is less critical, in that it is
only important in order for the
 training to be reasonably fast, but it does not impact on final result.
We choose for the first BP epoch $5\times 10^7$ training cycles
with learning rate $\eta=9\times 10^{-2}$ and momentum term
$\alpha=0.9 $, and for the second BP epoch 
$2.5\times 10^8$ training cycles
with learning rate $\eta=9\times 10^{-9}$ and momentum term
$\alpha=0.9 $.  

After these first two  training epochs, 
the diagonal $\chi^2_{\rm diag}=E_2$ per data point eq.~(\ref{chidiag}), 
which is being 
minimized, is of
order two for the central data set. This is comparable to the length
of training that was required to reach $\chi^2_{\diag}\approx 1.3$ for
the smaller data set of ref.~\cite{nn}.
The value of $E_3$ eq.~(\ref{chicov}), which is always bounded by it, 
$E_3 \le E_2 $ is accordingly smaller (see table~3).
The training algorithm then switches to GA minimization of
the $E_3$ eq.~(\ref{chicov}). The determination of the length of this
training epoch is critical, in that it controls the form of the final
fit. This can only be done by looking at the features of the full
Monte Carlo sample of neural networks.
\begin{table}[t]  
\begin{center}  
\begin{tabular}{|c||cc|cc|}
\multicolumn{5}{c}{TABLE 3} \\
\hline
 & \multicolumn{2}{c|}{A} & \multicolumn{2}{c|}{B} \\
\hline
 Experiment & $E_2$  &  $E_3$ 
& $E_2$  &  $E_3$ \\
\hline
Total & 2.05 & 1.54 & 2.03 & 1.36 \\
\hline
NMC       &  1.97 & 1.56 & 1.74  & 1.54 \\
BCDMS     &  1.93 & 1.66 & 1.32  & 1.26 \\
E665      &  1.64 & 1.37 & 1.83  & 1.38 \\
ZEUS94    &  3.15 & 2.26 & 3.01  & 2.21 \\
ZEUSBPC95 &  4.18 & 1.32 & 5.18  & 1.24 \\
ZEUSSVX95 &  3.37 & 1.88 & 5.68  & 2.11 \\
ZEUS97    &  2.33 & 1.54 & 3.02  & 1.37 \\
ZEUSBPT97 &  2.82 & 1.97 & 2.08  & 1.22 \\
H1SVX95   &  3.21 & 0.96 & 4.74  & 1.09  \\
H197      &  0.86 & 0.76 & 1.08  & 0.87  \\
H1LX97    &  1.96 & 1.46 & 1.50  & 1.18 \\
H199      &  1.15 & 1.07 & 1.10  & 1.01 \\
H100      &  1.59 & 1.50 & 1.48  & 1.26 \\
\hline
\end{tabular}
\caption{\small The uncorrelated 
$\chi^2$, $E_2$  eq.~(\ref{chidiag}) and  the total $\chi^2$, $E_3$  
eq.~(\ref{chicov}) for
  the fit to the central data points: (A)
after the backpropagation training epoch  and (B)
after the final genetic algorithms training epoch.}
\label{chis}
\end{center} 
\end{table}

Before addressing this issue, however, it turns out to be necessary to
consider the possibility of introducing cuts in the data set. Indeed,
consider the results
obtained after a GA training of $ 4\times 10^4$
cycles (with mutation rate $5\times 10^{-3}$) to the central data set,
displayed in
table~3. This is a rather long training: indeed, 
in each  GA training cycle all the data are shown
to the nets. Hence  in $ 4\times 10^4$ GA cycles the
data are shown to the nets $0.7\times10^8$ times, comparable to the
number of times they are shown to the nets during BP training.
It is apparent that whereas $E_3\sim1$ for most
experiments, it remains abnormally high for NMC and especially ZEUS94
and ZEUSSVX95. Because of the weighted training which has been
adopted, this is unlikely to be due to insufficient training of these
data sets, and is more likely related to problems of these data sets. 

Whereas ZEUSSVX95 only contains a small number of
data points, NMC and ZEUS94 account each for more than 10\% of the
total number of data points, and thus they can bias final results
considerably.
 The case of NMC was discussed in detail in
ref.~\cite{nn}.
This data set is the only one to cover the
medium-$x$, medium-small $Q^2$ region (compare figure~1) and thus it
cannot be excluded from the fit. As discussed in ref.~\cite{nn}, 
 the relatively large value of $E_3$ for this
experiment is a consequence of internal inconsistencies within
the data set. A
value of $E_3\approx1.5$ indicates that the neural nets do not
reproduce the subset of data which are incompatible with the
bulk, as it should be, whereas  a 
value $E_3\approx 1$ could only be obtained by
overlearning, i.e. essentially by fitting irrelevant fluctuations (see
ref.~\cite{nn}). 

Let us now consider the case of ZEUS94. The kinematic region of
this experiment is entirely covered by ZEUS97, H197, H199, H100. We
can therefore study the impact of excluding this experiment from the
global fit, without information loss. The results obtained in such
case are displayed in table~4: when the experiment is not fitted the
$E_3$ value for all experiments with which it overlaps improves
and so does the global $E_3$, whereas $E_3$
for ZEUS94 itself only deteriorates by a comparable amount, despite the fact
that the experiment is now not fitted at all. We conclude that the
experiment should be excluded from the fit, since its inclusion result
in a deterioration of the fit quality, whereas its exclusion does not entail
information loss. Difficulties in the inclusion of this experiment in
global fits were already pointed out in refs.~\cite{GKK,alekhin},
where it was suggested that they may be due to underestimated
or nongaussian uncertainties. 
It is likely that  ZEUSSVX95 has similar problems. However, its
inclusion 
in the fit is no
reason of concern, even if its high $E_3$ value were due to incompatibility
of this experiment with the others or underestimate of its
experimental uncertainties, because of the small number of
data points. It is therefore retained in the data set. Our final data
set thus includes all experiments in table~1, except ZEUS94.
We are thus left
with 1698 data points.
\begin{table}[t]  
\begin{center}  
\begin{tabular}{|cc|}
\multicolumn{2}{c}{TABLE 4} \\
\hline
 Experiment & $E_3$  \\
\hline
Total     & 1.25    \\
\hline
NMC       & 1.51    \\
BCDMS     & 1.24    \\
E665      & 1.23    \\
ZEUS94    & 2.28    \\
ZEUSBPC95 & 1.16    \\
ZEUSSVX95 & 2.08    \\
ZEUS97    & 1.37    \\
ZEUSBPT97 & 1.00    \\
H1SVX95   & 1.04    \\
H197      & 0.84    \\
H1LX97    & 1.19    \\
H199      & 1.00    \\
H100      & 1.24    \\
\hline
\end{tabular}
\caption{\small The same fit as the last column of table~3 if the ZEUS94
  data are excluded from the fit. }
\label{zeus94inc}
\end{center} 
\end{table}

For the sake of future applications, it is interesting to ask how the
procedure of selecting experiments in the data set can be
automatized. This can be done in an iterative way as follows: first, 
a neural net (or  sample of neural nets) is trained 
on only one experiment; then, 
the total $E_3$ for the full
data set is computed using this neural net (or sample of nets); the
procedure is then repeated for all experiments, and 
the experiment which leads to the smallest total $E_3$ is selected. In
the second iteration, the net (or sample of nets) is trained on the
experiment selected in the first iteration plus any of the other
experiments, thereby leading to the selection of a second experiment
to be added to that selected previously, and so on. The process can be
terminated before all experiments are selected, for instance if it is
seen that the addition of a new experiment does not lead to a
significant improvement in $E_3$ for given length of training.

\begin{figure}[t]
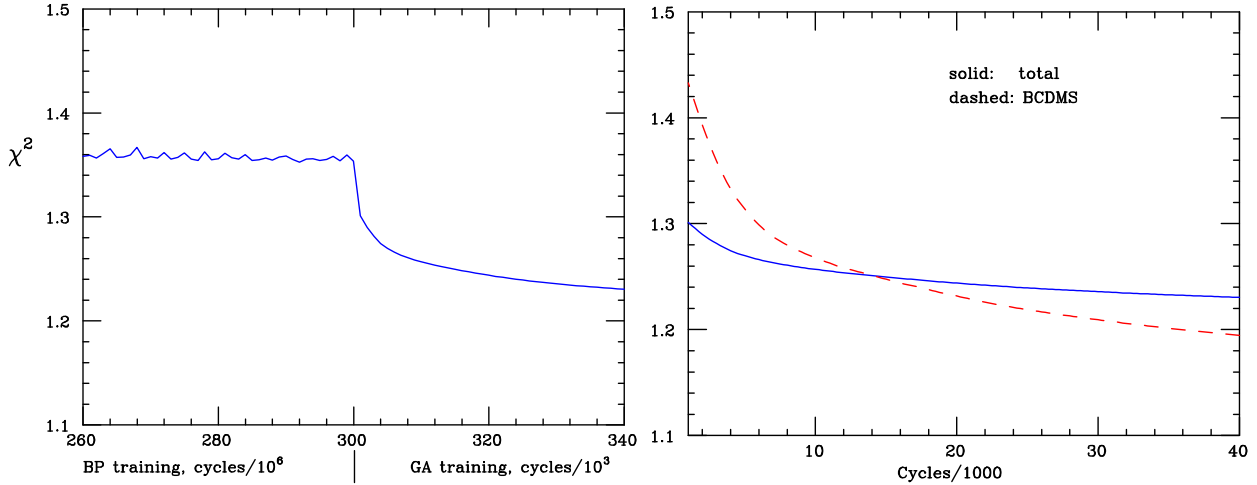

\begin{center}
\epsfig{width=0.495\textwidth,figure=chi2cme.ps}
\epsfig{width=0.47\textwidth,figure=chi2cme_bcd.ps}
\caption{\small Dependence of the $\chi^2$ eq.~(\ref{chi2}) on the length of
  training: (left) total training (right) detail of the GA training.}
\label{trlength}
\end{center}
\end{figure}
We now proceed to discuss the length of training for our final
 data set.
The $\chi^2$ eq.~(\ref{chi2}) is shown in figure~3 
as a function of the number of GA
 training cycles.
The $\chi^2$ decreases very rapidly during
the first few hundreds of training cycles.
After about 5000 training cycles, the $\chi^2$ as a function of the
 training length
essentially flattens for all experiments but BCDMS. The further
decrease of
 the total $\chi^2$ is then due essentially to the decrease  of the
 contribution from BCDMS. A
 training length of $ 4\times 10^4$ GA cycles is necessary in order for the
 $\chi^2$ of BCDMS to flatten out at $\chi^2\sim 1.2$. 
As discussed in ref.~\cite{nn}, the
 BCDMS data can only be learnt with a  longer training because they have  high
 precision while being located in the intermediate $x$ (valence)
 region, where the parton distribution displays significant variation.

 The $E_3$ values for the fit of a neural net to the
 central data with this training is given in table~3.  It
 shows that all experiments are well reproduced with the exceptions
discussed above. It is interesting to observe that while  $E_3$
 eq.~(\ref{chicov}) decreases significantly during the GA training,
 the uncorrelated  $E_2$ eq.~(\ref{chidiag}) decreases
 marginally, and in fact it actually increases for several HERA 
 experiments. This
 shows that correlations are sizable for the HERA experiments, so
 that the approach of ref.~\cite{nn}, based on the minimization of
 $\chi^2_{\diag}=E_2$ eq.~(\ref{chidiag}),
is not adequate in this case. GA minimization appears to be very
 efficient in reducing the $E_3$ value relatively fast.

We finally turn to the issue of theoretical constraints. The
only theoretical assumption on the shape of $\fd$ 
 is that it satisfies the kinematic constraint $F_2(1,Q^2)=0$
for all $Q^2$. As this constraint is local, its implementation
is straightforward: it can be enforced by including in the data set a
number
of artificial data points which satisfy the constraint with a suitably 
tuned error. In the present fit we have checked that
the best choice is to add a number of artificial  points at $x=1$,
equal to
2\% of the experimental trained points (33 points with ZEUS94 excluded
from the fits), and with error 
equal to one tenth of the mean statistical error of  the
trained points. These points are equally spaced in  $\ln Q^2$,
within the  range covered by the trained points. 

\section{Results}

The result of the minimization of a single neural net
to the central data points is shown
in table~4. The results for the final set of 1000 neural networks are  
displayed in table~5, while in
table~6 we give the details of results for each experiment. Note that
the figure of merit for the minimization $E_3$ eq.~(\ref{chicov}) and
its average defined in the appendix eq.~(\ref{avchi2}) differs from
the full $\chi^2$ eq.~(\ref{chi2}) not only because the latter is
computed from the structure function averaged over nets
eq.~(\ref{avf2}), but also because of the different treatment of
normalization errors in the respective covariance matrices
eq.~(\ref{covmatnn}) and eq.~(\ref{covmat}).

Besides the $\chi^2$ we
also list the values of various quantities, defined in the appendix,
which can be used to assess the goodness of fit.
\begin{table}[t]  
\begin{center}  
\begin{tabular}{|cc|} 
\multicolumn{2}{c}{
$F_2^p(x,Q^2)$}\\   
\hline
 $N_{\rep}$ & 1000 \\
\hline
$\chi^2$  & 1.18 \\
$\la E \ra$  & 2.52 \\
$r\lc F^{(\net)}\rc$ & 0.99  \\
${\cal R}$& 0.54  \\
\hline
$\la V\lc \sigma^{(\net)}\rc\ra_{\dat}$ & $1.2~10^{-3}$  \\
$\la PE\lc \sigma^{(\net)}\rc\ra_{\dat}$ &  $80\%$ \\

$\la \sigma^{(\mrexp)}\ra_{\dat}$ & 0.027   \\
$\la \sigma^{(\net)}\ra_{\dat}$ & 0.008   \\

$r\lc \sigma^{(\net)}\rc$ &  0.73   \\
\hline
$\la V\lc \rho^{(\net)}\rc\ra_{\dat}$ & 0.20 \\
$\la \rho^{(\mrexp)}\ra_{\dat}$ &  0.31\\
$\la \rho^{(\net)}\ra_{\dat}$ & 0.67 \\
$r\lc \rho^{(\net)}\rc$ &  0.54 \\
\hline
$\la V\lc \mathrm{cov}^{(\art)}\rc\ra_{\dat}$ & $3.3~10^{-7}$ \\
$\la \mathrm{cov}^{(\mrexp)}\ra_{\dat}$ & $1.3~10^{-4}$ \\
$\la \mathrm{cov}^{(\net)}\ra_{\dat}$ & $3.6~10^{-5}$ \\
$r\lc \mathrm{cov}^{(\net)}\rc$ & 0.49 \\
\hline
\end{tabular}
\caption{\small Estimators of the final  results.}
\label{results_new}
\end{center} 
\end{table}

\begin{table}[t]  
\begin{center}
\tiny
\begin{tabular}{|c|ccc|}    
\hline
 Experiment & NMC & BCDMS & E665 \\
\hline
$\chi^2$  & 1.47  & 1.19 & 1.20 \\
$\la E \ra $ & 2.69  & 3.17 & 2.29 \\
$r\lc F^{(\net)}\rc$ &  0.96  & 0.99 & 0.91 \\
${\cal R}$&  0.59 & 0.50 & 0.56 \\
\hline
$\la V\lc \sigma^{(\net)}\rc\ra_{\dat}$ & $0.002$ & $1.9~10^{-5}$ 
& 0.0013 \\
$\la PE\lc \sigma^{(\net)}\rc\ra_{\dat}$ & 0.63 & 0.56 & 0.89 \\

$\la \sigma^{(\mrexp)}\ra_{\dat}$ & 0.017 & 0.007 & 0.032 \\
$\la \sigma^{(\net)}\ra_{\dat}$ & 0.008  & 0.005 & 0.008 \\

$r\lc \sigma^{(\net)}\rc$ &  0.23   & 0.98 & 0.17 \\
\hline
$\la V\lc \rho^{(\net)}\rc\ra_{\dat}$ & 0.51  & 0.69 & 0.29 \\

$\la \rho^{(\mrexp)}\ra_{\dat}$ & 0.17  & 0.52 & 0.20\\
$\la \rho^{(\net)}\ra_{\dat}$ & 0.84  & 0.86 & 0.60 \\
$r\lc \rho^{(\net)}\rc$ & 0.08 & 0.73 & 0.05 \\
\hline
$\la V\lc \mathrm{cov}^{(\art)}\rc\ra_{\dat}$ & $2.4~10^{-9}$ &
$1.8~10^{-9}$  &  $4.5~10^{-9}$\\
$\la \mathrm{cov}^{(\mrexp)}\ra_{\dat}$ &  $4.4~10^{-5}$ & 
$3.8~10^{-5}$&  $1.7~10^{-4}$ \\
$\la \mathrm{cov}^{(\net)}\ra_{\dat}$  &  $ 5.2~10^{-5}$ &
 $2.3~10^{-5}$&  $3.3~10^{-5}$ \\
$r\lc \mathrm{cov}^{(\net)}\rc$ & -0.03 & 0.98 & 0.16 \\
\hline
\end{tabular}\\\bigskip
\begin{tabular}{|c|ccccccccc|}    
\hline
 Experiment &  ZEUSBPC95 & ZEUSSVX95 &
ZEUS97 & ZEUSBPT97 & H1SVX95 & H197 & H1LX97 & H199 & H100 \\
\hline
$\chi^2$  & 1.02 & 2.08 & 1.35 & 
0.86& 0.67 & 0.71& 1.07 & 0.90 & 1.11 \\
$\la E \ra $ & 2.07 & 2.03 & 2.24 
& 2.08 & 2.03 & 1.91 & 2.41 & 1.93 & 2.11 \\
$r\lc F^{(\net)}\rc$ & 0.98 & 0.96 & 0.99& 0.99&
0.97 & 0.99 & 0.99& 0.98 & 0.99 \\
${\cal R}$& 0.51 & 0.66 & 0.55& 0.55&
0.44 & 0.46 & 0.53& 0.48 & 0.54 \\
\hline
$\la V\lc \sigma^{(\net)}\rc\ra_{\dat}$ 
& $4.3~10^{-4}$ & 0.0035 & 0.0010 &$1.3~10^{-4}$& 0.0043& 
0.0030 & 0.0005& 0.003& 0.0013 \\

$\la PE\lc \sigma^{(\net)}\rc\ra_{\dat}$ & 0.91 &
0.94&0.87& 0.72& 0.96& 0.95 & 0.75 & 0.96 & 0.93\\

$\la \sigma^{(\mrexp)}\ra_{\dat}$ & 0.022 & 0.061 &
0.037& 0.012 & 0.063& 0.040 & 0.027 & 0.051 & 0.030 \\
$\la \sigma^{(\net)}\ra_{\dat}$ & 0.006 & 0.013
& 0.011 & 0.006 & 0.011 & 0.008& 0.008& 0.008 & 0.009 \\

$r\lc \sigma^{(\net)}\rc$ & 0.85 &0.72 & 0.86&
0.73 & 0.84 & 0.87 & 0.42 & 0.82 & 0.89\\
\hline
$\la V\lc \rho^{(\net)}\rc\ra_{\dat}$ & 0.09 &
0.30 & 0.12 & 0.14& 0.118 & 0.14 & 0.31& 0.16 & 0.14\\

$\la \rho^{(\mrexp)}\ra_{\dat}$ & 0.61 & 0.24 &
0.28 & 0.40 & 0.36 & 0.06 & 0.29& 0.05 & 0.09\\
$\la \rho^{(\net)}\ra_{\dat}$ & 0.77 & 0.64 &
0.39& 0.63 & 0.57 & 0.27 &0.58& 0.29 & 0.26 \\
$r\lc \rho^{(\net)}\rc$ & 0.53 & 0.40 &
0.66& 0.60 & 0.48 & 0.51 &0.69 & 0.37 & 0.55 \\
\hline
$\la V\lc \mathrm{cov}^{(\art)}\rc\ra_{\dat}$ & $6.4~10^{-8}$& $1.9~10^{-6}$ &
$3.4~10^{-7}$& $1.4~10^{-9}$& $3.0~10^{-6}$&
$3.8~10^{-7}$& $3.8~10^{-8}$& $2.7~10^{-7}$& $1.7~10^{-7}$\\

$\la \mathrm{cov}^{(\mrexp)}\ra_{\dat}$ & $2.8~10^{-4}$& $8.5~10^{-4}$&
$3.7~10^{-4}$& $5.8~10^{-5}$ &
0.0014 & $1.0~10^{-4}$& $2.1~10^{-4}$ & 
$1.4~10^{-4}$& $9.6~10^{-5}$\\

$\la \mathrm{cov}^{(\net)}\ra_{\dat}$  & $2.8~10^{-5}$ & $1.2~10^{-4}$&
$3.2~10^{-5}$& $2.3~10^{-5}$ & $7.0~10^{-5}$&
$1.510^{-5}$& $6.9~10^{-5}$ & $1.6~10^{-5}$& 
$2.2~10^{-5}$\\

$r\lc \mathrm{cov}^{(\net)}\rc$ & 0.69 & 0.48 & 
0.77 & 0.65 & 0.53 & 0.61& 0.57 & 0.54 & 0.58\\
\hline
\end{tabular}
\caption{\small Final results for the individual 
experiments: fixed target (top) and HERA (bottom) }
\end{center} 
\label{resultsexp_new}
\end{table}

The quality of the final fit is somewhat better than that of the fit
to the central data points shown in table~4. In particular, with the
exception of NMC (which is likely to have internal inconsistencies~\cite{nn}) and
ZEUSSVX95 (which is likely to have the same problems as those of ZEUS94
discussed in section~4) the $\chi^2$ per degree of freedom is of order
1 for all experiments. It is interesting to note that the $\chi^2$ for
the neural network average is rather better than the average
$\langle E_3 \rangle$ eq.~(\ref{avchi2}).  The (scatter)
correlation between experimental data and the neural network
prediction equals one to about 1\% accuracy, with the exception of NMC,
ZEUSSVX95 (which have the aforementioned problems) and E665. The
E665 kinematic region
overlaps almost entirely (apart from very small $Q^2<1$~GeV$^2$) with
that of NMC and BCDMS, while having lower accuracy (this is why the
experiment was not included in the fits of ref.~\cite{nn}). The data
points corresponding to this experiment are therefore essentially
predicted by the fit to other experiments, thus explaining the
somewhat smaller scatter correlation.

The  average neural network variance is in general substantially smaller than
the average experimental error, typically by a factor $3-4$. This is
the reason why $\la E \ra > \chi^2$: the
neural nets fluctuate less about central experimental values than the
Monte Carlo replicas. In the presence of substantial
error reduction, the (scatter) correlation between
network covariance and experimental error is generally not very high,
and can take low values when a small number of data points from one
experiment is enough to determine the outcome of the fit, such as in
the case of the NMC experiment, even more so for E665.~\cite{nn}

\begin{figure}[t]
\begin{center}
\epsfig{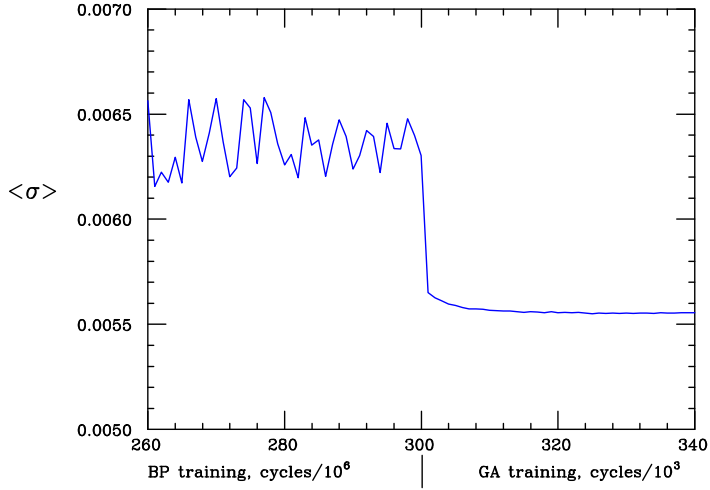}
\caption{\small Dependence of $\la\sigma^{(\net)}\ra_{\dat}$ on the length of
  training for the BCDMS experiment.}
\label{bcdsig}
\end{center}
\end{figure}
 As
discussed extensively in ref.~\cite{nn} it is important to make sure
that this is due to the fact that information from individual
data points is combined through an underlying law by the neural
networks, and not due to parametrization bias. To this purpose, the ${\cal
  R}$-estimator has been introduced in
ref.~\cite{nn}, where it was shown that in the presence of substantial
error reduction ${\cal
  R}\gsim 1$ if there is parametrization bias, whereas ${\cal
  R}\approx 0.5$ in the absence of parametrization 
bias.\footnote{Note that in ref.~\cite{nn} the $\cal R$-ratio was defined in terms of
the diagonal $\chi^2_{\diag}$ eq.~(\ref{chidiag}), 
because neural networks were trained by
minimizing that quantity. It is easy to see that the results of
ref.~\cite{nn} for $\cal R$ remain true when the full $E_3$
eq.~(\ref{chicov}) is
minimized, provided $\cal R$ is redefined accordingly.} It is apparent
from tables~5-6 that indeed ${\cal
  R}\approx 0.5$ for all experiments. Note that, contrary to what was
found in ref.~\cite{nn}, there is now some error reduction also for the
BCDMS experiment, though by a somewhat smaller  amount than for other
experiments. We will come back to this issue  when comparing
results to those of ref.~\cite{nn}. 

Further evidence that the error reduction is not due to
parametrization bias can be obtained by studying the dependence of
$\la\sigma^{(\net)}\ra_{\dat}$ on the length of training. This
dependence 
is shown in fig.~\ref{bcdsig} for the BCDMS experiment. It is apparent that the
error reduction is correlated with the goodness of fit displayed in
fig.~\ref{trlength}, and it occurs during the GA training, thereby suggesting that
error reduction occurs when the neural networks start reproducing an
underlying law. If error reduction were
due to parametrization bias it would be essentially independent of the
length of training. 

\begin{figure}[t]
\begin{center}
\epsfig{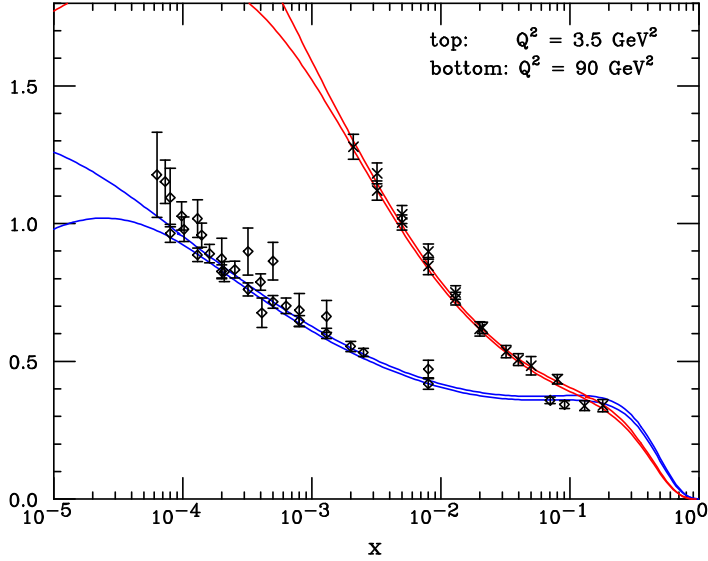}
\caption{\small Final results for $\fd$ compared to data. For the
 neural net result, the
 one-$\sigma$ error band is shown.}
\label{f2plot}
\end{center}
\end{figure}
The point-to-point correlation $\rho$ of the neural nets is somewhat
larger than that of the data, as one might expect as a consequence of
an underlying law which is being learnt by the neural nets. 
In fact, for the NMC experiment the increase in correlation
essentially compensates the reduction in error, in such a way that the
average covariance of the nets and the data are essentially the
same. This again shows that in the case of the NMC
experiment a small number of points is sufficient to predict the
remaining ones.
 For  all other experiments, however, the
covariance of the nets is substantially smaller than that of the
data. As a consequence the (scatter) correlation of covariance
remains relatively high for all experiments, except NMC, and
especially E665 whose points are essentially predicted by the fit to
other experiments.

\begin{figure}[t]
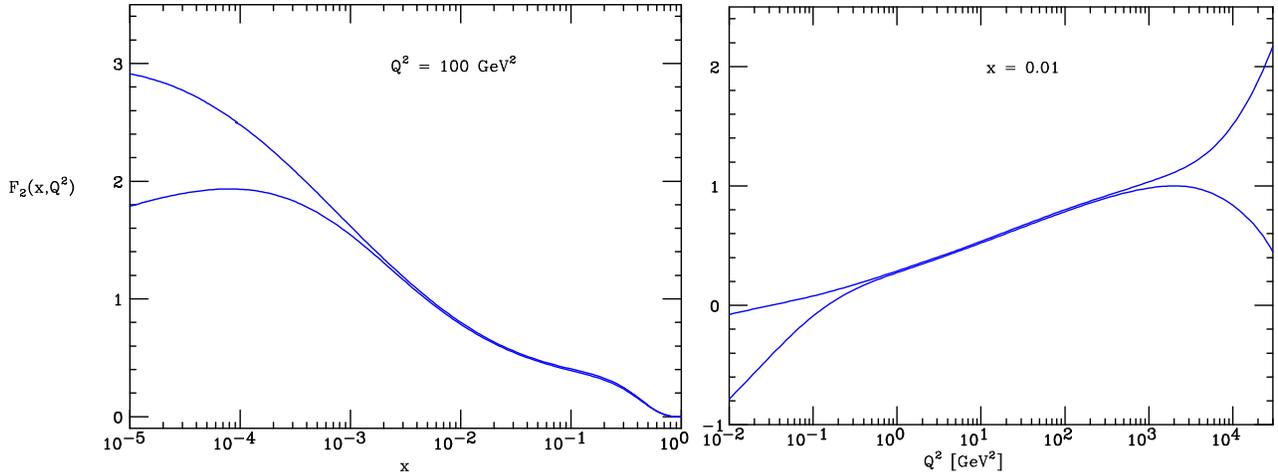

\begin{center}
\epsfig{width=0.535\textwidth,figure=f2_extra_x.ps}
\epsfig{width=0.45\textwidth,figure=f2_extra_q2.ps}
\caption{\small One-$\sigma$ error band for the structure function
  $F_2(x,Q^2)$ 
computed from neural
  nets. Note the different scale on the $y$ axis in the two plots.}
\label{f2extra}
\end{center}
\end{figure}
The structure function and associated one-$\sigma$ error band is
compared to the data as a function of $x$ for a pair of typical
values of $Q^2$ in fig.~\ref{f2plot}. In fig.~\ref{f2extra} the behaviour 
of the structure
function as a function of $x$ at fixed $Q^2$ and as a function of
$Q^2$ at fixed $x$ is also shown. It is apparent that in the data
region the error on the neural nets is rather smaller than that on the
data used to train them. The error however grows rapidly as soon as
the nets are extrapolated outside the region of the data. At large
$x$, however, the
extrapolation is kept under control by the kinematic constraint
$F_2(1,Q^2)=0$. 

\begin{figure}[t]
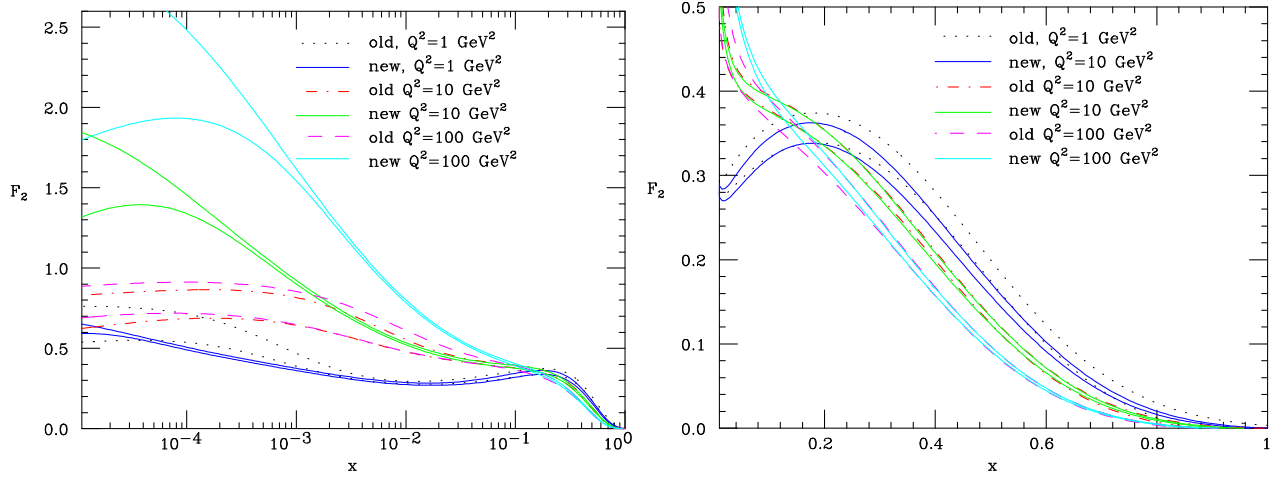

\begin{center}
\epsfig{width=0.49\textwidth,figure=f2_x_log.ps}
\epsfig{width=0.49\textwidth,figure=f2_x_lin.ps}
\caption{\small Comparison of the parametrization of $F_2(x,Q^2)$ of
  ref.~\cite{nn} (old) with that of the present paper (new). The pairs
  of curves correspond to a one-$\sigma$ error band.}
\label{f2comp}
\end{center}
\end{figure}
\begin{figure}[t]
\begin{center}
\epsfig{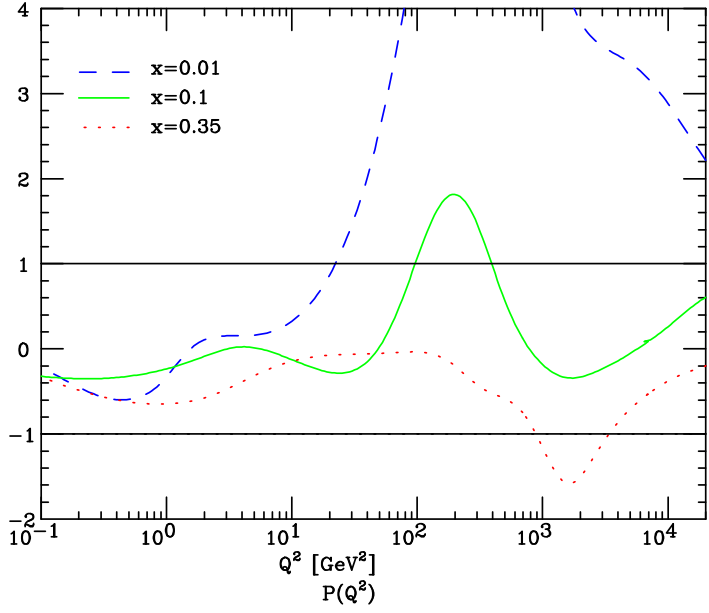}
\caption{\small The relative pull eq.~(\ref{pull}) of the new and old $F_2$
  parametrizations. The one-$\sigma$ band is also shown.} 
\label{pullfig}
\end{center}
\end{figure}

Let us finally compare the determination of $F_2(x,Q^2)$ presented
here with that of ref.~\cite{nn}, which was based on pre-HERA data. 
In fig.~\ref{f2comp} one-$\sigma$ error bands for the two
parametrizations are compared, whereas in fig.~\ref{pullfig} we
display the relative pull of the two parametrizations, defined as
\be
P(x,Q^2)\equiv \frac{F_2^{new}(x,Q^2)-F_2^{old}(x,Q^2)}
{\sqrt{\sigma_{old}^2(x,Q^2)+\sigma_{new}^2(x,Q^2)}},
\label{pull}
\ee
where $\sigma(x,Q^2)$ is the error on the structure function
determined as the variance of the neural network sample.
In view of the fact that the old fit only included BCDMS and NMC data,
it is interesting to consider four regions:  (a)
 the BCDMS region (large $x$, intermediate $Q^2$, e.g. $x=0.3$,
 $Q^2=20$~GeV$^2$); (b)
the NMC region (intermediate $x$, not too large $Q^2$, e.g. $x=0.1$,
$Q^2=2$~GeV$^2$); (c) the HERA region  
(small $x$ and large $Q^2$, e.g. $0.01$ and $Q^2>10$~GeV$^2$); (d) the region
where neither the old nor the new fit had data (very large or very
small $Q^2$). In region (a) the new fit is rather more precise than
the old one, for reasons to be discussed shortly, while central values
agree, with $P\lsim 1$). In region (b) the new fit is significantly more
precise than the old one, while central values agree to about one
sigma. In region (c) the new fit is rather accurate while
the old fit had large errors, but $P\gg1$ nevertheless, 
because the HERA rise of $F_2$ is outside the error bands extrapolated
from NMC. This shows that even though errors on  extrapolated data
grow rapidly they become unreliable when extrapolating far from the data.
Finally in region (d) all errors are very large and $P$ is
consequently small, except at small $x$ and large $Q^2$, where the new
fits extrapolate the rise in the HERA data, which is missing
altogether in the old fits.

Let us finally come to the issue of the BCDMS error, which, as
already mentioned, is reduced somewhat in the current
fit in comparison to the data and the previous fit. This may appear 
surprizing, in that the new fit does not contain any new data in
the BCDMS region.
However,
 as is apparent
from fig.~\ref{bcdsig}, this error reduction takes place in the GA
training stage, when $E_3$ eq.~(\ref{chicov}) is
minimized. Furthermore, we have verified that if the uncorrelated
$\chi^2_{\diag}=E_2$ eq.~(\ref{chidiag}) is minimized during the GA training
no error reduction is observed for BCDMS. 
Hence, we conclude that the reason why error reduction for BCDMS  was
not found in
ref.~\cite{nn} is that in that reference neural networks were trained by 
minimizing $E_2$. In fact, as discussed in sect.~4, 
the BCDMS experiment turns out to require the longest time to learn, 
especially after inclusion of the HERA data. Error reduction only
obtains after this lenghty minimization process.

\section{Conclusion}
We have presented a determination of the probability density in the
space of structure functions for the structure function $\fd$ 
for the proton, based on all available data from the  NMC,
BCDMS,
E665, ZEUS and H1 collaborations. Our results take the form of a Monte
Carlo sample of 1000 neural networks, each of which provides a
determination of the structure function for all $(x,Q^2)$. The
structure function and its  statistical moments 
(errors, correlations and so on) can be determined by
averaging over this sample. Results are made available as a
FORTRAN routine which gives $F_2(x,Q^2)$, determined by a  set of
parameters, and 1000 sets of parameters corresponding to the Monte
Carlo sample of structure functions. They can be downloaded
from the web site {\tt
http://sophia.ecm.ub.es/f2neural/}. 

This works updates and upgrades that of ref.~\cite{nn}, where similar
results were obtained from the BCDMS and NMC
data only. The main improvements in the present work are related to the
need of handling a large number of experimental data, 
affected by large correlated systematics. 
Apart from many
smaller technical aspects, the main improvement introduced here is the
use of genetic algorithms to train neural networks on top of
back-propagation. This has allowed for a more accurate handling of
correlated systematics.

Whereas the results of this paper may be of direct practical use for
any application where an accurate determination of $F_2^p(x,Q^2)$ and
its associate error are necessary, its main motivation is the
development of a set of techniques which will be required for the
construction of a full set of parton distributions with faithful
uncertainty estimation based on the same method. This will be
presented in a forthcoming publication.

\vskip 1cm

{\large {\bf Acknowledgments}} 

\vskip 3mm
We thank M.~Arneodo, E.~Rizvi, E.~Tassi and F.~Zomer for informations
on the HERA data, and
  G.~d'Agostini, L.~Garrido and G.~Ridolfi for discussions.
This work
has been supported by the project 
GC2001SGR-00065  and by the Spanish
 grant AP2002-2415.

\vfill
\eject   

\appendix

\section{Statistical estimators}
\label{estimators}
We define various statistical estimators which have been used in the
paper.
The superscripts $(\dat)$, $(\art)$ and $(\net)$ refer respectively to
the original data, to the $N_{\rep}$ Monte Carlo replicas of the data,
and to the  $N_{\rep}$  neural networks.
The subscripts $\rep$ and $\dat$ refer respectively to whether averages 
are taken  by
summing over all replicas or over all data.

\begin{itemize}
\item{\bf Replica averages}
\begin{itemize}
\item Average over the number of replicas for each experimental point
  $i$
\be
\la
 F_i^{(\art)}\ra_{\rep}=\frac{1}{N_{\rep}}\sum_{k=1}^{N_{\rep}}
F_i^{(\art)(k)}\ .
\ee 
\item Associated variance
\be
\label{var}
\sigma_i^{(\art)}=\sqrt{\la\lp F_i^{(\art)}\rp^2\ra_{\rep}-
\la F_i^{(\art)}\ra^2_{\rep}} \ .
\ee
\item Associated covariance
\be
\label{ro}
\rho_{ij}^{(\art)}=\frac{\la F_i^{(\art)}F_j^{(\art)}\ra_{\rep}-
\la F_i^{(\art)}\ra_{\rep}\la F_j^{(\art)}\ra_{\rep}}{\sigma_i^{(\art)}
\sigma_j^{(\art)}} \ .
\ee
\be
\label{cov}
\mathrm{cov}_{ij}^{(\art)}=\rho_{ij}^{(\art)}\sigma_i^{(\art)}
\sigma_j^{(\art)} \ .
\ee
\item Mean variance and percentage error on central values 
over the
  $N_{\dat}$ data points.
\be
\la V\lc\la F^{(\art)}\ra_{\rep}\rc\ra_{\dat}=
\frac{1}{N_{\dat}}\sum_{i=1}^{N_{\dat}}\lp \la F_i^{(\art)}\ra_{\rep}-
F_i^{(\mrexp)}\rp^2 \,
\ee
\be
\la PE\lc\la F^{(\art)}\ra_{\rep}\rc\ra_{\dat}=
\frac{1}{N_{\dat}}\sum_{i=1}^{N_{\dat}}\lc\frac{ \la F_i^{(\art)}\ra_{\rep}-
F_i^{(\mrexp)}}{F_i^{(\mrexp)}}\rc \ .
\ee
We define analogously $\la V\lc\la
\sigma^{(\art)}\ra_{\rep}\rc\ra_{\dat} $, $\la V\lc\la
\rho^{(\art)}\ra_{\rep}\rc\ra_{\dat}$, $\la V\lc\la
\mathrm{cov}^{(\art)}\ra_{\rep}\rc\ra_{\dat}$ and $\la PE\lc\la
\sigma^{(\art)}\ra_{\rep}\rc\ra_{\dat}$. 
\item Scatter correlation:
\be
r\lc F^{(\art)}\rc=\frac{\la F^{(\mrexp)}\la F^{(\art)}
\ra_{\rep}\ra_{\dat}-\la F^{(\mrexp)}\ra_{\dat}\la\la F^{(\art)}
\ra_{\rep}\ra_{\dat}}{\sigma_s^{(\mrexp)}\sigma_s^{(\art)}} \,
\ee
where the scatter variances are defined as
\be
\sigma_s^{(\mrexp)}=\sqrt{\la \lp F^{(\exp)}\rp^2\ra_{\dat}-
\lp \la  F^{(\exp)}\ra_{\dat}\rp^2} \,
\ee
\be
\sigma_s^{(\art)}=\sqrt{\la \lp \la F^{(\art)}\ra_{\rep}\rp^2\ra_{\dat}-
\lp \la  \la F^{(\art)}\ra_{\rep} \ra_{\dat}\rp^2} \ .
\ee 
We define analogously $r\lc\sigma^{(\art)}\rc$, $r\lc\rho^{(\art)}\rc$
and  $r\lc\mathrm{cov}^{(\art)}\rc$. Note that the scatter correlation
and
scatter variance are not related to the variance and correlation
Eqs. \ref{var}-\ref{cov}.
\item Average variance:
\be
\la \sigma^{(\art)}\ra_{\dat}=\frac{1}{N_{\dat}}
\sum_{i=1}^{N_{\dat}}\sigma_i^{(\art)} \ .
\label{avvar}
\ee
We  define analogously $\la\rho^{(\art)}\ra_{\dat}$ and
$\la\mathrm{cov}^{(\art)}\ra_{\dat}$,  as well as the
corresponding experimental quantities.
\end{itemize}
\item{\bf Neural network averages}
\begin{itemize}
\item Average $\chi^2$ over nets
\be
\la E
\ra=\frac{1}{N_{\rep}}\sum_{k=1}^{N_{\rep}}{E_3}^{(k)} \ ,
\label{avchi2}
\ee
where $E_3^{(k)}$ is the $\chi_2$ given in eq.~(\ref{chicov}).
\item Mean variance and percentage error on central values over the
  $N_{\dat}$ data points.
\be
\la V\lc\la F^{(\net)}\ra_{\rep}\rc\ra_{\dat}=
\frac{1}{N_{\dat}}\sum_{i=1}^{N_{\dat}}\lp \la F_i^{(\net)}\ra_{\rep}-
F_i^{(\mrexp)}\rp^2 \,
\label{vnets}
\ee
\be
\la PE\lc\la F^{(\net)}\ra_{\rep}\rc\ra_{\dat}=
\frac{1}{N_{\dat}}\sum_{i=1}^{N_{\dat}}\lc\frac{ \la F_i^{(\net)}\ra_{\rep}-
F_i^{(\mrexp)}}{F_i^{(\mrexp)}}\rc \ .
\label{penets}
\ee
\item
We define analogously percentage errors on the correlation and
covariance. 
\item Scatter correlation
\be
r\lc F^{(\net)}\rc=\frac{\la F^{(\mrexp)}\la F^{(\net)}
\ra_{\rep}\ra_{\dat}-\la F^{(\mrexp)}\ra_{\dat}\la\la F^{(\net)}
\ra_{\rep}\ra_{\dat}}{\sigma_s^{(\mrexp)}\sigma_s^{(\net)}} \ .
\label{sccnets}
\ee 
We define analogously $\la\rho^{(\net)}\ra_{\dat}$ and
$\la\mathrm{cov}^{(\net)}\ra_{\dat}$. 
\item  ${\cal R}$-ratio
\be
{\cal R}=\frac{\langle \tilde E \rangle}{\langle  E \rangle},
\label{rratdef}
\ee
where ${\langle  E \rangle}$ is given by eq.~(\ref{avchi2}) and 
\bea
\la \widetilde{E}
\ra&=&\frac{1}{N_{\rep}}\sum_{k=1}^{N_{\rep}}{\widetilde{E}}^{(k)}
\label{avchi2til}\\
\label{tilchicov}
{\widetilde{E}}^{(k)}&=&\frac{1}{N_{\dat}}\sum_{i,j=1}^{N_{\dat}}\lp
F_i^{(\net)(k)}-F_i^{(\exp)(k)}\rp{\overline\mathrm{cov}^{(k)}}^{-1}_{ij}
\lp
F_j^{(\net)(k)}-F_j^{(\exp)(k)}\rp .
\eea

\end{itemize}
\end{itemize}

\end{document}